\documentclass[prd,preprint,tightenlines,floatfix,showpacs,preprintnumbers,nofootinbib,eqsecnum,superscriptaddress]{revtex4}

 \usepackage[dvips,final]{graphicx}
  \usepackage{amssymb}
   \usepackage{amsmath}
    \usepackage{amsfonts}
     \usepackage{epsfig}
      \usepackage{bm}

\begin{document}

\title{Exclusive diffractive photoproduction of dileptons by timelike 
Compton scattering}

\author{W.~Sch\"afer}
\email{Wolfgang.Schafer@ifj.edu.pl}
\affiliation{Institute of Nuclear Physics PAN, PL-31-342 Cracow,
Poland} 
\author{G.~\'Slipek}
\email{Gabriela.Slipek@ifj.edu.pl}
\affiliation{Institute of Nuclear Physics PAN, PL-31-342 Cracow,
Poland} 
\author{A.~Szczurek}
\email{Antoni.Szczurek@ifj.edu.pl}
\affiliation{Institute of Nuclear Physics PAN, PL-31-342 Cracow,
Poland} 
\affiliation{University of Rzesz\'ow, PL-35-959 Rzesz\'ow,
Poland}

\date{\today}

\begin{abstract}
We derive the forward photoproduction amplitude for 
the diffractive $\gamma p \to l^+ l^- p$ reaction 
in the momentum space.
within the formalism of $k_\perp$- factorization.
Predictions for the $\gamma p \to l^+ l^- p$ reaction 
are given using unintegrated gluon distribution from 
the literature.
We calculate the total cross section as a function of 
photon-proton center of mass energy and the invariant mass 
distribution of the lepton pair.
We also discuss whether the production of timelike virtual 
photons can be approximated by continuing to the spacelike 
domain $ q^2 < 0$.
The present calculation provides an input for future 
predictions for exclusive hadroproduction in 
the $p\, p \to p\, \, l^+ l^- p$ reaction.
\end{abstract}


\maketitle

\section{Introduction}

\begin{figure}[!ht]    %
\begin{center}
\includegraphics[width=0.4\textwidth]{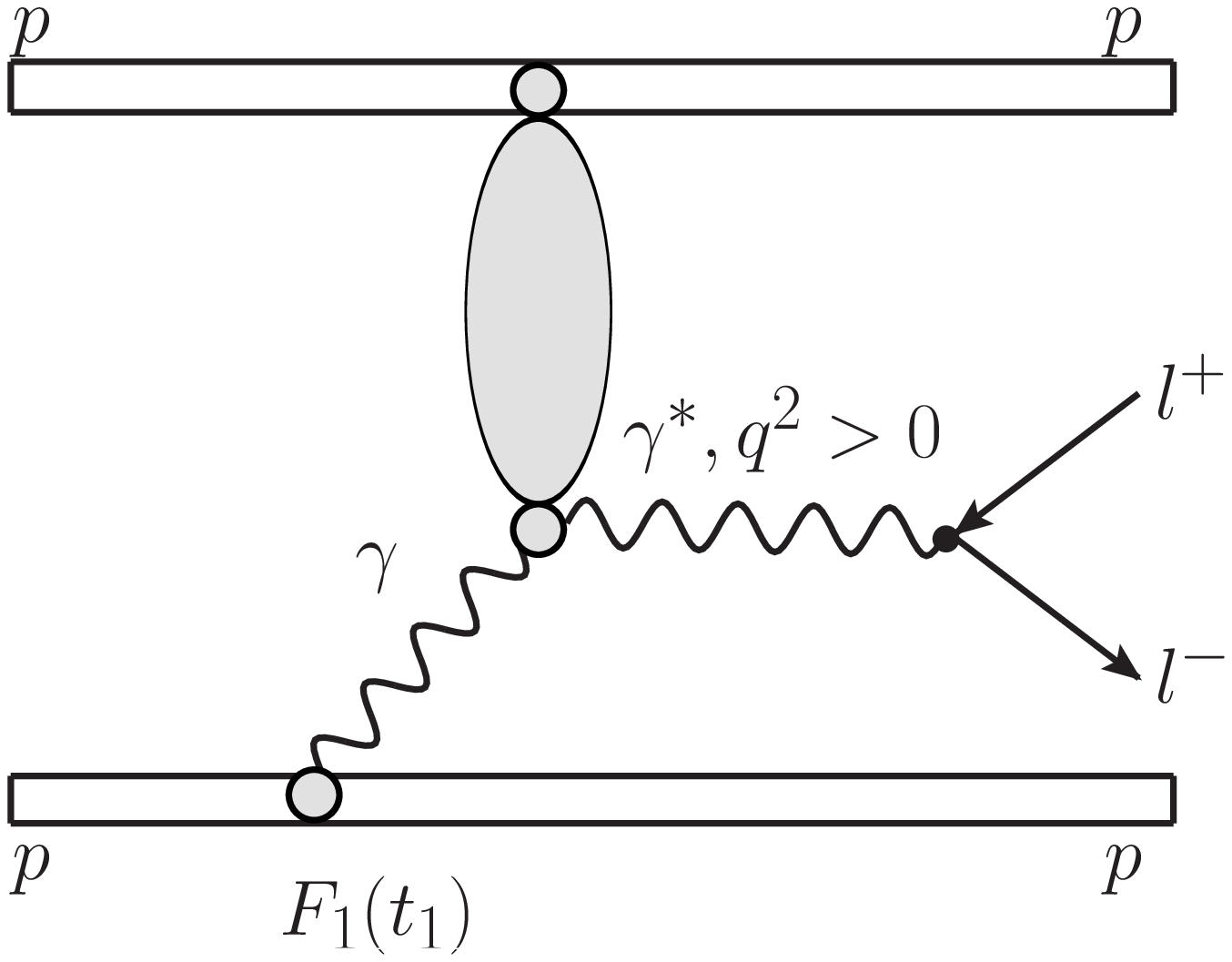}
\hspace{0.5cm}
\includegraphics[width=0.4\textwidth]{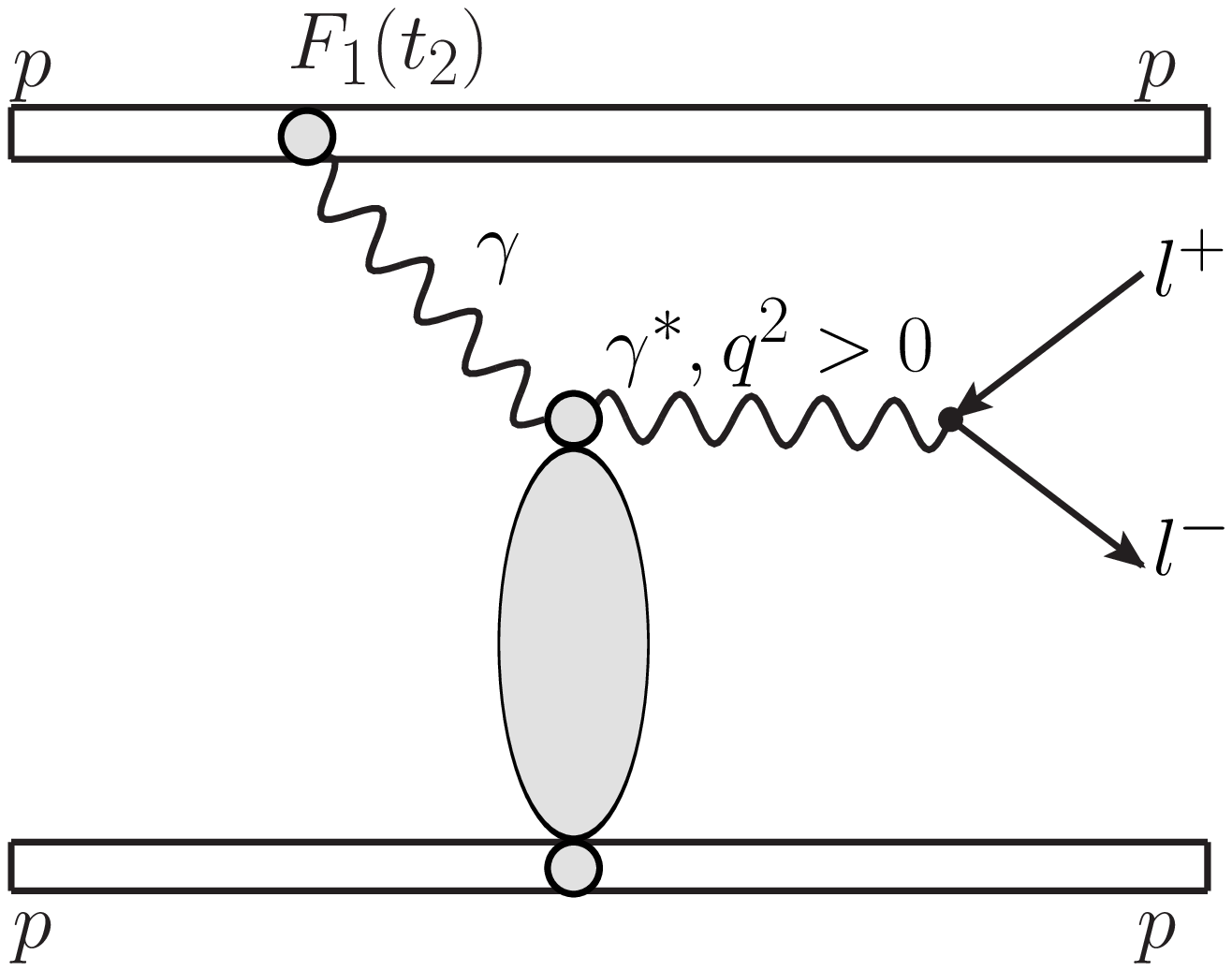}
\end{center}
   \caption{\label{fig:diagrams_pp_to_epempp}
   \small
An example of the non-QED mechanism for the prodution of opposite charge 
leptons in the $p p \to p p l^+ l^-$ reaction. 
}
\end{figure}

Measuring absolutely normalized cross sections at the LHC
is of great importance for the high-energy physics community. 
This requires having a well understood luminosity monitor.
Following the pioneering work \cite{Budnev}, the QED 
process $pp \to p l^+ l^- p$ via photon--photon fusion is often discussed
as a process which can be used for measuring the luminosity at the LHC
\cite{T02,P97,KCS08}.
It is therefore very important to estimate other non-QED contributions to 
exclusive $l^+ l^-$ production.
One possible source of dileptons is the exclusive production of vector mesons
or $Z$--bosons (see e.g. \cite{SS07,RSS08,CSS09}).
The dilepton pairs originating from these processes 
however have invariant masses 
close to the mass of the decaying state.
In Figure \ref{fig:diagrams_pp_to_epempp} we show an exclusive diffractive 
mechanism which produces continuum dilepton pairs, 
and hence may compete with the standard QED process. 
In this reaction the coupling of the photon to the proton is known 
and can be expressed in terms of the nucleon electromagnetic form factors. 
At the small transferred momenta ($t_1$  or $t_2$) of relevance, it is sufficient, 
in the high energy limit, to include the Dirac electromagnetic form factor.
Besides its role as a possible background to electromagnetic lepton pair production,
the $\gamma p \to \gamma^* p$ amplitude may contain interesting 
information on the small--$x$ gluon
distribution in the nucleon.

In the present work we shall concentrate on the photon-pomeron
subprocess. In Fig.\ref{fig:diagram_gammap_to_epemp} we show a QCD
mechanism, where  
the photon splits into a quark-antiquark pair which interacts
with the proton through the exchange of an off-diagonal QCD gluon ladder. 
In principle this process could have been studied at HERA. 
In Fig.\ref{fig:diagram_gammap_to_epemp} 
the incoming photon is spacelike, (or quasireal) but the outgoing photon is 
timelike, i.e. its virtuality $q^2 >0$.
This process is often called timelike Compton scattering (TCS) in the literature,
although the specific mechanism considered by us is maybe better termed
a QCD version of (virtual) Delbr\"uck scattering.
A collinear factorisation treatment of timelike Compton scattering 
in terms of the nucleon's skewed
(mainly quark-) distributions can be found in \cite{Berger,Belitsky}.
This approach is most relevant for lower center-of-mass energies.
We will restrict ourselves to high energies, where the $t$-channel
exchange is dominated by gluons, and choose a $k_\perp$--factorization formalism
very similar to the one used in diffractive vector meson production
\cite{I03,INS06}.  

The TCS cross section has also been evaluated
in a color-dipole model with a saturation-idea inspired dipole-nucleon
cross section \cite{M08}. However there both incoming and outgoing photons 
were assumed to be spacelike. A recent estimate of high energy cross--sections
in leading order collinear factorization, without explicit gluons, 
is found in \cite{PSW09}.

In this work we present the momentum--space formulation of timelike Compton
scattering at small-$x$, taking due account of the timelike nature of the
final state photon.

This paper is organized as follows.
In the next section, we present the formalism which is used in our calculations.
In Sec. III we present the main results for the $\gamma p \to l^+ l^- p$ reaction.
Finally, in the last section we summarize our results and show further perspectives.

\section{Formalism}


\begin{figure}[!ht]    %
\begin{center}
\includegraphics[width=0.6\textwidth]{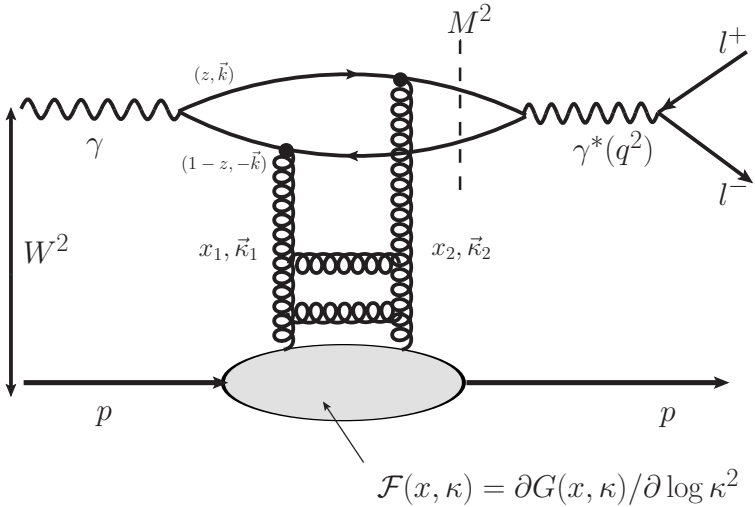}
\end{center}
   \caption{\label{fig:diagram_gammap_to_epemp}
   \small
The diagram for the production of virtual timelike photons.
}
\end{figure}

\vspace{0.3cm}

The photoproduction amplitude will be the major building block for 
our prediction of exclusive dilepton pair production.
The amplitude for the reaction is shown schematically in 
Fig.\ref{fig:diagram_gammap_to_epemp}.
In the diagram, we distinguish three stages of the process:
first the incoming real photon fluctuates into a quark-antiquark pair,
then a gluon ladder is exchanged  between the $ q \bar q$ pair and the proton
and finally the $ q \bar q$ pair recombines to form a virtual photon
which subsequently decays into a lepton--antilepton pair.

The amplitude of the subprocess $\gamma p \to \gamma^*(q^2) p$ 
is a sum of the contributions for a given flavour $f$ of quarks in the loop.
\begin{eqnarray}
{\cal M}(\gamma p \to \gamma^*(q^2) p)  = \sum \limits_{f} {\cal M}_f
(\gamma p \to \gamma^*(q^2) p).
\end{eqnarray}
Here the initial state photon is real, and hence transversely polarized.
We take only the dominant $s$--channel helicity conserving contribution
into account, and suppress helicities of photons/protons in our notation.
The calculation of the amplitude follows the same procedure as for the exclusive 
production of vector mesons, which is explained in great detail in Ivanov's thesis 
\cite{I03}. The main difference is that the final state light--cone wavefunction is 
replaced by a free quark propagator times the QED--spinor structure for the
$q \bar q \to \gamma^*$ transition.
The forward $\gamma p \to \gamma^* p$ amplitude for a given flavour contribution can 
then be written as:
\begin{eqnarray}
{\cal M}_f(\gamma p \to \gamma^*(q^2) p)  = W^2\,  4 \pi \alpha_{\mathrm{em}} \, e_{f}^2 \, 2 \, \int_{0}^{1} dz\,\int_{0}^{\infty} \pi dk_{\bot}^2{{\cal A}_{f}(z,k_{\bot}^2,W^2)\over [k_{\bot}^2+m_f^2-z(1-z)q^2-i\varepsilon]}
\nonumber \\
\nonumber \\
= W^2\, 4 \pi \alpha_{\mathrm{em}} \,  e_{f}^2\, 2\, \cdot 2\,  \int_{0}^{1/2} \frac{dz}{z(1-z)} \,
\int_{0}^{\infty} \pi dk_{\bot}^2 { {\cal A}_{f}(z,k_{\bot}^2,W^2) \over 
\left [{k_{\bot}^2+m_f^2\over z(1-z)}\,- q^2-i\varepsilon \right ] },
\label{eq:amplitude}
\end{eqnarray}
where the explicit form of ${\cal A}_{f}(z,k_{\bot}^2,W^2)$ will be discussed below,
$\alpha_{\mathrm{em}}$ is the QED fine--structure constant; $e_f =\frac{2}{3}$ for $u, c, t$ and
$e_f = -\frac{1}{3} $ for $d, s, b$ is the quark charge.
The transverse momentum squared of (anti-)quarks is denoted by $k^2$, 
their longitudinal momentum fractions are $z$ and $1-z$, respectively.
The running coupling $\alpha_s$ enters at the scale $q^2 = {\mathrm{max}}
\{\kappa^2,k^2 + m_f^2 \}$,
where $m_f$ is the quark mass for flavor $f$.
Now notice, that the invariant mass of the $q \bar q$ pair is given by 
\begin{eqnarray}
M^2 = {k_{\bot}^2 + m_f^2\over z(1-z)} \, ,
\end{eqnarray}
so that the second line of Eq.(\ref{eq:amplitude}) suggests a change of variables from 
$(z,k_{\bot}^2) \to (M^2, k_{\bot}^2)$:
\begin{equation} 
{dz \, dk_{\bot}^2 \over z(1-z)} \longrightarrow {dM^2 \over M^2} \, {dk_{\bot}^2 \over J}
\label{eq:variable_transformation} 
\end{equation}
Using the $z \leftrightarrow 1-z$ symmetry we could restrict ourselves to $0 \leq z \leq 1/2$,
so that from
\begin{eqnarray}
z = z(M^2,k_{\bot}^2) = {1\over 2}\left [1-\sqrt{1-4\frac{k_{\bot}^2+m_f^2}{M^2}} \, \right],
\end{eqnarray}
we obtain the jacobian factor
\begin{eqnarray}
J = \sqrt{1-4 \left (\frac{k_{\bot}^2+m^2}{M^2} \right)},
\end{eqnarray}
which introduces an integrable singularity (see e.g. Eq. \ref{eq:variable_transformation}).
The integration domain is transformed as
\begin{equation}
\{ 0 \leq z \leq 1/2 \} \times \{ 0 \leq k_{\bot}^2 < \infty \} \longrightarrow  \{ 4m_f^2 \leq M^2 < \infty\}
\times \{ 0 \leq k_{\bot}^2 \leq (M^2 - 4m_f^2)/4 \}. 
\end{equation}
Finally, we can cast the amplitude in the form 
\begin{eqnarray}
{\cal M}_f(\gamma p \to \gamma^*(q^2) p) = W^2\, 16 \pi^2 \alpha_{\mathrm{em}} e_{f}^2\ \cdot \int_{4m_f^2}^{\infty} \,{a_{f}(W^2,M^2)\over M^2-q^2-i\varepsilon}\, dM^2.
\label{eq:integral}
\end{eqnarray}
Here $a_{f}(W^2,M^2)$ is related to the diffractive amplitude for the 
$\gamma p \to q \bar q p$ transition \cite{Nikolaev:1991et}, 
however with the spinorial contractions from the final 
state performed. For the lack of a better name we will refer to it as the spectral distribution
or spectral density. 
Its imaginary part is given by the integral:
\begin{eqnarray}
\Im m \, a_{f}(W^2,M^2) &=& {1 \over M^2} \int_{0}^{{1 \over 4}M^2-m_f^2}\, \frac{dk_{\bot}^2}{J}\,
\Im m \, {\cal A}_{f}(z(M^2,k_{\bot}^2),k_{\bot}^2,W^2) \, .
\end{eqnarray}
with 
\begin{eqnarray}
\Im m {\cal A}_{f}(z,k_{\bot}^2,W^2) &&= \pi \int_{0}^{\infty} {\pi d\kappa^2  \over \kappa^4} 
\alpha_s(q^2) {\cal F}(x_{\mathrm{eff}},\kappa^2) \Big[ A_{0f}(z,k_{\bot}^2) \, W_{0f}(k_{\bot}^2,\kappa^2) 
\nonumber \\
&& + A_{1f}(z,k_{\bot}^2)\,W_{1f}(k_{\bot}^2,\kappa^2) \Big] \, ,
\end{eqnarray}
where the auxiliary functions $A_{1,0 f},W_{1,0 f}$ can be taken from Ref. \cite{RSS08}:
\begin{eqnarray}
A_{0f}(z,k_{\bot}^2) &=& m_f^2, \,  \, 
A_{1f}(z,k_{\bot}^2) = [z^2+(1-z)^2]{k_{\bot}^2 \over k_{\bot}^2+m_f^2},
\nonumber \\
W_{0f}(k_{\bot}^2,\kappa^2) &=& {1 \over {k_{\bot}^2+m_f^2}} - {1 \over \sqrt{(k_{\bot}^2 - m_f^2 - \kappa^2)^2 + 4m_f^2 k_{\bot}^2 }},
\nonumber \\
W_{1f}(k_{\bot}^2,\kappa^2) &=& 1- {k_{\bot}^2+m_f^2 \over 2k_{\bot}^2} \left(1 + {k_{\bot}^2- m_f^2 - \kappa^2 \over  \sqrt{(k_{\bot}^2 - m_f^2 - \kappa^2)^2 + 4m_f^2 k_{\bot}^2 }}\right).
\end{eqnarray}
Given the relation to the diffractive $\gamma p \to q \bar q p$ amplitude we feel justified
to obtain the real part from the standard derivative form of the dispersion relation:
\begin{equation}
\Re e \,  {\cal A}_{f}(z,k_{\bot}^2,W^2)  = \Im m  {\cal A}_{f}(z,k_{\bot}^2,W^2) \cdot
\tan\Big( {\pi \over 2} {\partial \log \Im m  {\cal A}_{f}(z,k_{\bot}^2,W^2) \over 
\partial \log W^2} \Big).
\end{equation}
The function ${\cal F}(x,\kappa^2)$ is the unintegrated gluon distribution of the proton, which at 
large values of gluon transverse momenta can be expressed in terms of the collinear gluon distribution 
as
\begin{eqnarray}
{\cal F}(x,\kappa^2) = { \partial G (x,\kappa^2)  \over \partial \log \kappa^2 },
\end{eqnarray} 
In the present analysis we use an unintegrated gluon distribution obtained by 
fitting to the structure function data measured at HERA \cite{IN02}.
Following \cite{I03,INS06}, we correct for skewedness effects by taking the unintegrated 
at $x_{\mathrm{eff}} = c_{\mathrm{skewed}} \, {M^2 \over W^2}$ \,with $c_{\mathrm{skewed}} = 0.41$.
To calculate the integral in Eq.(\ref{eq:integral}) for $q^2 > 4m_{f}^2$ we use the 
Plemelj-Sokhocki formula:
\begin{eqnarray}
{1\over x - i\varepsilon}
= \mathrm{PV} {1 \over x} + i \pi \delta(x),
\label{eq:PS_formula}
\end{eqnarray}
where $\mathrm{PV}$ denotes the principal value integral.
We can finally represent real and imaginary part of the forward TCS
amplitude as
\begin{eqnarray}
\Im m  {\cal M}_f(\gamma p \to \gamma^*(q^2) p) &&= W^2\, 16 \pi^2 \alpha_{\mathrm{em}} e_{f}^2
\cdot \Big\{
\theta(4m_f^2 - q^2) \,  \int_{4m_f^2}^{\infty}\, dM^2 {\Im m \,a_{f}(W^2,M^2)\over M^2-q^2}
\nonumber \\
&&+ \theta(q^2 - 4m_f^2) \, 
\Big( \mathrm{PV} 
\int_{4m_f^2}^\infty dM^2 {\Im m\,a_{f}(W^2,M^2) \over M^2 - q^2}
\, + \pi \, \Re e \,a_{f}(W^2,q^2) 
\Big)
\Big\},
\nonumber \\
\Re e  {\cal M}_f(\gamma p \to \gamma^*(q^2) p) &&= W^2\, 16 \pi^2 \alpha_{\mathrm{em}} c_{f}^2
\cdot \Big\{
\theta(4m_f^2 - q^2) \,  \int_{4m_f^2}^{\infty}\, dM^2 {\Re e \,a_{f}(W^2,M^2)\over M^2-q^2}
\nonumber \\
&&+ \theta(q^2 - 4m_f^2) \, 
\Big( \mathrm{PV} 
\int_{4m_f^2}^\infty dM^2 {\Re e\,a_{f}(W^2,M^2) \over M^2 - q^2}
\, - \pi \, \Im m \,a_{f}(W^2,q^2) 
\Big)
\Big\}.
\nonumber \\
\label{eq:final}
\end{eqnarray}
Two comments are in order on the final form of the TCS amplitude (\ref{eq:final}):
firstly, the ``$i \pi$''--terms from the decomposition (\ref{eq:PS_formula})
lead to a quite nontrivial structure of the amplitude in terms of the spectral 
distribution $a_f$. In particular, the TCS amplitude will not be a monotonous
function of $q^2$. Secondly, we should remember that these  
``$i \pi$''--terms derive from the cut through the $q \bar q$ pair which exists
in the perturbative amplitude, when $q^2 > 4m_f^2$. Clearly in the real hadronic world,
there are no cuts due to quarks going on--shell, and our amplitude must, as usual,
be interpreted in a parton--hadron duality sense.

As a last step we must extend our amplitude to finite momentum transfers. For simplicity
we will assume it to have the following factorized form
\begin{equation}
{\cal M}_f(\gamma p \to \gamma^*(q^2) p;t) =  {\cal M}_f(\gamma p \to \gamma^*(q^2) p) \, \exp[Bt] \, ,
\end{equation}
which should be sufficiently accurate for small $t$, within the diffraction cone.
The total cross section for the $\gamma p \to \gamma^* p$ process can then 
be obtained as
\begin{eqnarray}
\sigma (\gamma p \to \gamma^*(q^2) p) =  { (\Re e { {\cal M} \over W^2} )^2 + (\Im m {{\cal M} \over W^2}  )^2  \over 16 \pi B},
\end{eqnarray}
where for a first estimation we shall take $B = 4\, \mathrm{GeV}^{-2}$.

The invariant mass distribution of dileptons for the $\gamma p \to l^+ l^- p$
reaction which can be accessed in experiment is given by:

\begin{eqnarray}
{d \sigma \over d q^2}(\gamma p \to l^+l^-p) = 
{\alpha_{\mathrm{em}} \over 3 \pi q^2} \cdot \sigma (\gamma p \to \gamma^*(q^2) p),
\end{eqnarray}
where $l^+ l^-$ means either $e^+ e^-$ or $\mu^+ \mu^-$.
This simple formula applies only when $q^2 \gg m_{l}^2$.

\section{Results}

\begin{figure}[!ht]    %
\begin{center}
\includegraphics[width=0.4\textwidth]{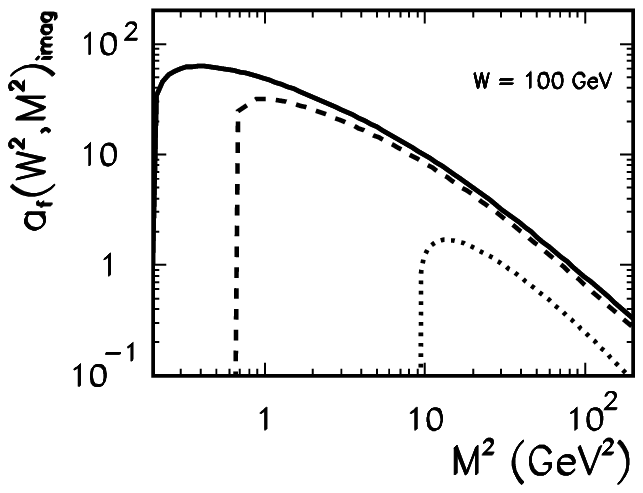}
\includegraphics[width=0.4\textwidth]{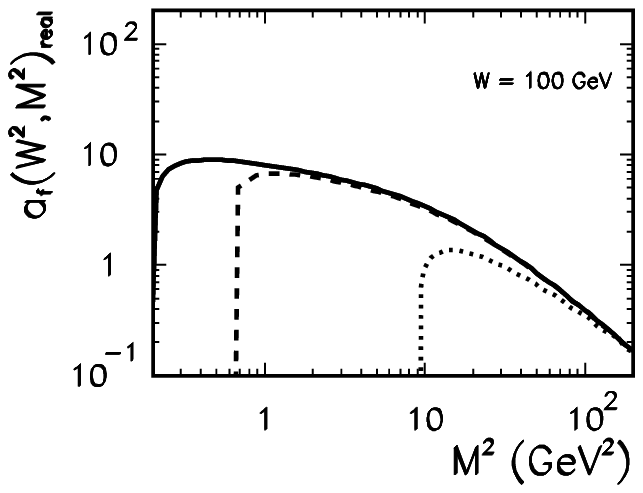}\\
\includegraphics[width=0.4\textwidth]{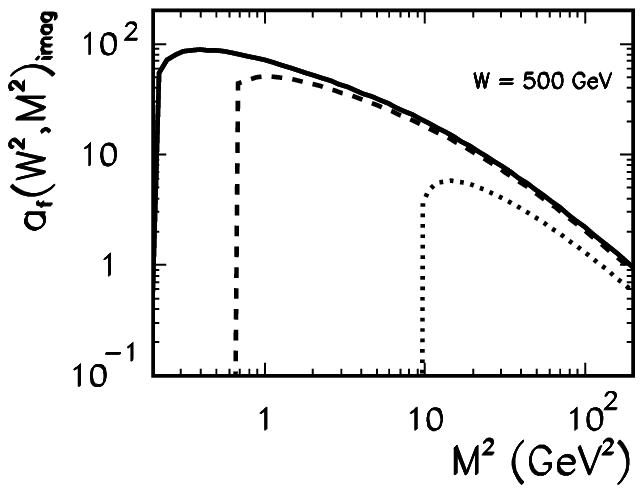}
\includegraphics[width=0.4\textwidth]{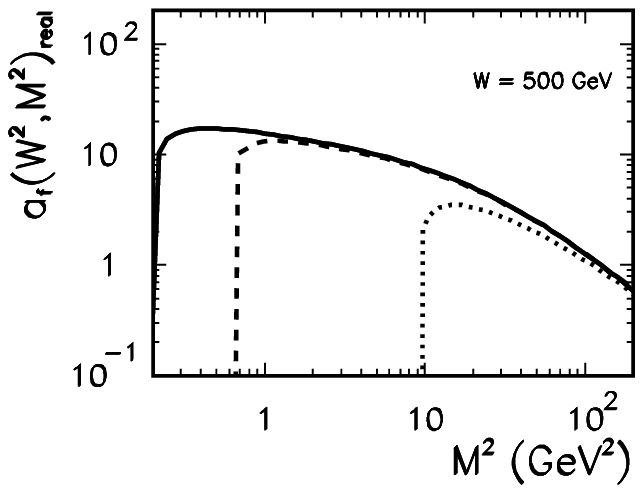}\\
\includegraphics[width=0.4\textwidth]{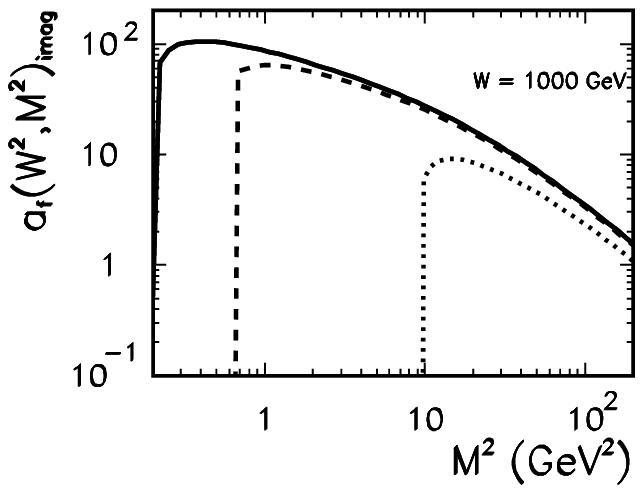}
\includegraphics[width=0.4\textwidth]{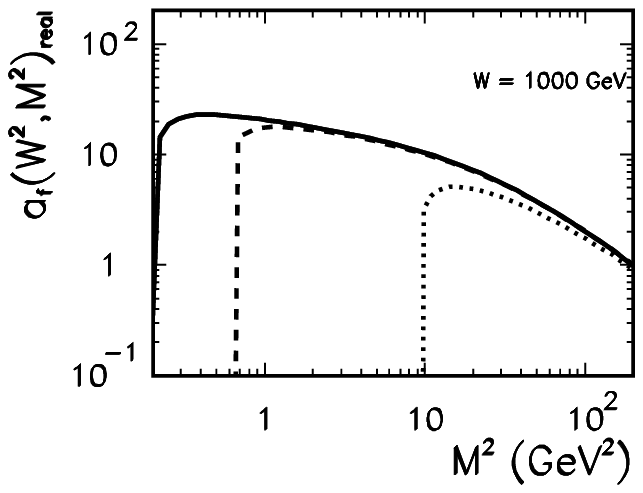}
\end{center}
   \caption{\label{fig:spectral_density_log}
   \small
Spectral density $(a_f)$ for different flavours: solid line for u,d, 
dashed line for s, 
dotted line for c at W =100, 500, 1000 GeV.
In the left row we show the imaginary part $\Im m \,a_{f}$, 
while in the right row, we show the real part $\Re e \,a_{f}$.
}
\end{figure}

Here we present predictions for the $\gamma p \to l^+ l^- p$ reaction.
In Fig.\ref {fig:spectral_density_log} we show spectral density 
as a function of $M^2$ (invariant mass of the $q \bar q$ pair)
for different $\gamma p$ center of mass energies: $ W = 100, 500, 1000$ GeV 
and for $u, d, s, c$ quarks separately.
The energy dependence of the spectral density derives from the $x_{\mathrm{eff}}$--
dependence of the unintegrated gluon density. 
This is why we observe a growth of the spectral density with energy.
For large invariant mass of the $q \bar q$ pair the spectral density 
tends to zero, which ensures the convergence of the integrals in (\ref{eq:final}).


\begin{figure}[!ht]    %
\begin{center}
\includegraphics[width=0.6\textwidth]{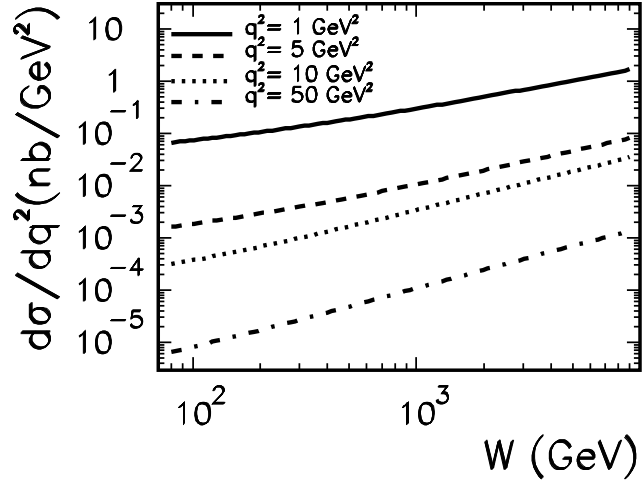}
\end{center}
   \caption{\label{fig:sigma_w}
   \small
The cross section for  $\gamma p \to l^+ l^- p$ as a function of center-of-mass photon-proton
energy for fixed values of the dilepton invariant mass.
}
\end{figure}

In Fig. \ref{fig:sigma_w} we show the invariant mass distribution
$d\sigma/dq^2$
of dileptons in the  $\gamma p \to l^+ l^- p$ reaction as a function of photon-proton
center-of-mass energy at fixed values of the invariant 
masses of the dilepton pairs ( $q^2$). 
Here all flavours are included in the amplitude. 
In general, the higher invariant mass, the faster 
growth with the photon-proton energy. This points to the fact
that the unintegrated glue is probed at on average harder scales,
where it has a faster $x$--dependence. 


\begin{figure}[!ht]    %
\begin{center}
\includegraphics[width=0.6\textwidth]{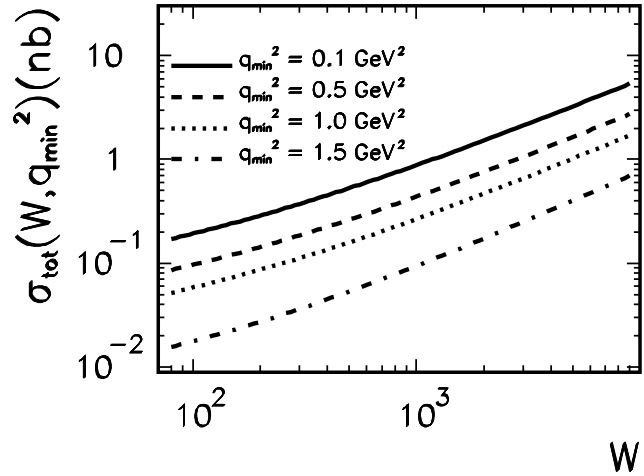}
\end{center}
   \caption{\label{fig:sigma_w_q2_min}
   \small
${\sigma}_{tot} (\gamma p \to l^+ l^- p ; q_{min}^2)$ for $q_{min}^2 = 0.1, 0.5, 1.0, 1.5 \,GeV^2$.
}
\end{figure}
In Fig. \ref{fig:sigma_w_q2_min} we show cross section integrated over dilepton invariant mass
\begin{eqnarray}
{\sigma}_{tot} (\gamma p \to l^+ l^- p ; q_{min}^2) = 
\int_{q_{min}^2}^{\infty}{d\sigma \over dq^2} dq^2 \, .
\label{integ_over_q2}
\end{eqnarray}
These cross sections are by a factor of about 5 larger than those in \cite{M08}.

\begin{figure}[!ht]    %
\begin{center}
\includegraphics[width=0.6\textwidth]{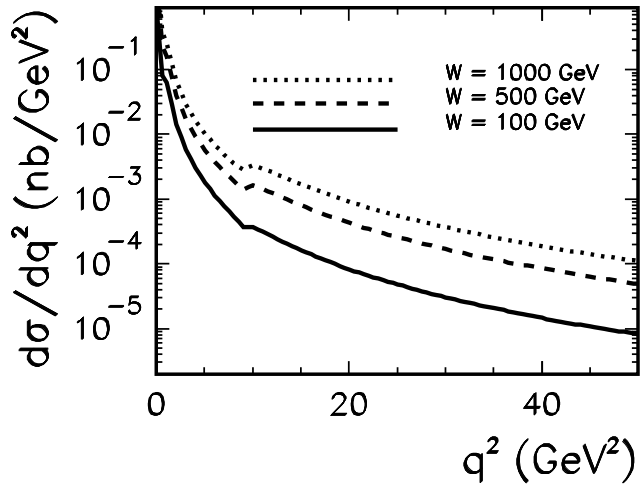}
\end{center}
   \caption{\label{fig:dsig_dq2}
   \small
Distribution in $q^2$  for W = 100, 500, 1000 GeV. 
}
\end{figure}

In Fig.\ref{fig:dsig_dq2} we show the invariant mass distribution of dileptons 
$d\sigma/dq^2$ as a function of dilepton invariant mass $q^2$ at fixed values of
$\gamma p$ energy.
An interesting feature of the invariant mass distribution is a cusp
at  $q^2 \sim  9 \, {\mathrm{GeV}}^2$. Notice that this is
the vicinity of $q^2 \sim 4 m_c^2$. Indeed the cusp is caused by 
the $c \bar c$ contribution to the amplitude which changes sign in this region.
This is a unique feature of the structure of the amplitude (\ref{eq:final})
with timelike final state photons. Here one should however remember the caveat
on the absence of quark thresholds, most optimistically one may
hope that such a structure survives in the vicinity of $q^2 \sim 4m_D^2$.
\begin{figure}[!ht]    %
\begin{center}
\includegraphics[width=0.6\textwidth]{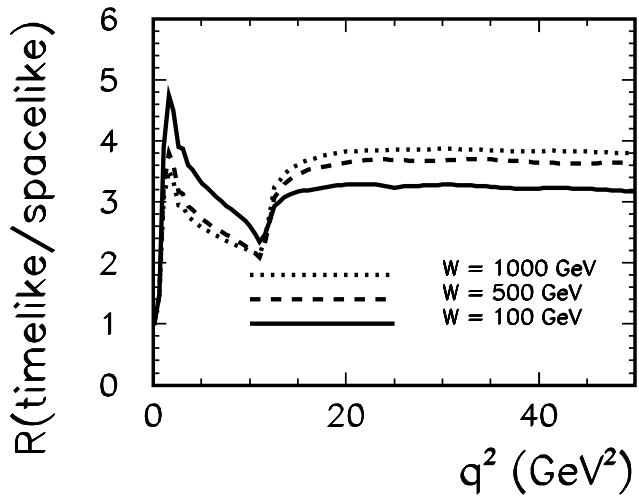}
\end{center}
   \caption{\label{fig:time-spacelike}
   \small
The ratio of the cross section 
${{d\sigma \over dq^2}timelike \over {d\sigma \over dq^2}spacelike}$ 
as a function of $q^2$.
}
\end{figure}
Finally, it is interesting to investigate how well the dilepton mass spectrum
can be calculated from the amplitude for production of spacelike 
photons in the final state.
In this case one would replace Eq.(\ref{eq:final}) by the straightforward
\begin{eqnarray}
{\cal{M}}_f^{{\mathrm{spacelike}}}(\gamma p \to \gamma^*(q^2)  p) 
= W^2\, 16 \pi^2 \alpha_{\mathrm{em}} e_{f}^2
\cdot \int_{4m_f^2}^{\infty} \,{ a_f(W^2,M^2)\over M^2+q^2}\, dM^2.
\label{spacelike}
\end{eqnarray}
Notice that in distinction to the timelike case, 
this is a monotonous function of $q^2$.
In Fig. \ref{fig:time-spacelike} we show the ratio of 
$d \sigma /dq^2$ for the (correct) timelike photons and 
the spacelike prescription as a function of $q^2$ at various fixed
energies. 
We observe that, as expected, the spacelike prescription does not
reproduce the structure present in the timelike amplitude.
At large $q^2$ it gives a fairly reasonable description of
the $q^2$--dependence, but is not able to reproduce the correct timelike
results. 
The cross section for timelike photons is bigger 
by a factor of  3-4 compared to the spacelike photon prescription.
\section{Conclusions}

We have derived the amplitude for the exclusive diffractive
photoproduction of lepton pairs in the $k_\perp$-factorisation
approach in the momentum space. We have discussed several
details of the formalism as well as differences compared to the
existing calculation in the literature which ignored 
the fact that the "produced" photons are timelike.

We have calculated the cross sections as a function
of photon-proton center of mass energy as well as a function of 
dilepton invariant mass. We have demonstrated how important is
the inclusion of correct dynamics 
(timelike outgoing photons instead of spacelike outgoing photons).
As a consequence the cross sections obtained 
here are significantly larger than those obtained
in the literature. We furthermore observed an 
interesting structure in the invariant mass distribution of
dileptons around $q^2 \sim 9 \,\mathrm{GeV}^2$.

The amplitude for the $\gamma p \to l^+ l^- p$ is
the main ingredient of the
diffractive amplitude for the $p p \to l^+ l^- p p$ process.
In the case of hadroproduction of dileptons the diffractive
mechanism constitutes a background to the purely electromagnetic
(photon--photon fusion) production of dileptons. The latter process
was sugested in the literature as a luminosity monitor
the LHC studies. How important is this background
for differential distributions in the four-body $p p \to  l^+ l^- pp$ 
reaction will be studied in detail elsewhere.

\vspace{1cm}

{\bf Acknowledments}

We are indebted to Janusz Chwastowski and Krzysztof Piotrzkowski for an interesting
discussion. This work was partially supported by the
MNiSW grants: N N202 236937, N N202 249235 and N N202 191634.


\end{document}